# Exploring the Influence of Dynamic Disorder on Transport Gap in Solid Pentacene

Zhiping Wang,[1,†] Sahar Sharifzadeh,[1,‡] Zhenfei Liu[1,¶], Peter Doak[1,§] and Jeffrey B. Neaton[1,2,3]

[1]Molecular Foundry, Lawrence Berkeley National Laboratory, Berkeley, California 94720, United States
[2]Department of Physics, University of California Berkeley, Berkeley, California 94720, United States
[3]Kavli Energy NanoScience Institute at Berkeley, Berkeley, California 94720, United States

**Abstract**: We combine a GW approach and *ab initio* Molecular Dynamics (AIMD) simulations to study the impact of thermal effects on transport gap in solid pentacene ($C_{22}H_{14}$). The dynamic disorder induced by thermal fluctuations is simulated by AIMD, providing the ensemble-averaged density of states (DOS) near the band gap. The GW corrected DOS, averaged over hundreds of snapshots from AIMD simulation containing disordered structures indicates that the edge-to-edge transport gap is 2.1±0.04 eV, reduced by ~0.1 eV in contrast to the static 0 K GW calculation. The peak-to-peak gap is found to be 2.7eV in excellent agreement with experiment after corrections for the surface and the Frank-Condon effects and providing fully *ab initio* agreement with experiment where previous theory required *ad hoc* Gaussian broadening and temperature corrections.

[†] Present address: School of Physics, Shandong University, Jinan 250100, P. R. China
[‡] Present address: Department of Electrical and Computer Engineering, Boston University, Boston, MA 02215, United States
[¶]Present address: Department of Chemistry, Wayne State University, Detroit, MI 48202, United States
[§] Present address: Center for Nanophase Material Science, Oakridge National Laboratory, Oakridge Tennessee 37831, United States

# 1. Introduction

Organic semiconductors are promising next generation energy conversion materials with many advantages over their inorganic counterparts, such as low-cost solution-phase chemical synthesis and device flexibility[1–6]. However, these materials are usually sensitive to temperature[7,8] because the molecules are held together by weak intermolecular interactions. Thermal fluctuations at room temperature are sufficient to break the translational symmetry of the lattice[9], which will consequently change the band structure as well as the charge transport properties.

To enhance the application of organic semiconductors in photovoltaic devices, it is crucial to determine the fundamental quantities such as transport and optical band gaps precisely with the thermal effect taken into account. Pentacene is a well-studied organic semiconductor with relevance to energy conversion, and has been the focus of many recent theoretical[10–14] and experimental[15–21] studies. In a previous work[14], the spectroscopic and transport properties in solid pentacene were systematically studied with many-body perturbation theory within the GW approximation and the Bethe-Salpeter equation approach. However finite temperature effects were accounted for only with a Gaussian broadening of the GW density of states justified by an estimate of thermal disorder. Using this approximation the GW quasiparticle transport gap agreed quantitatively with experimental results[18,19] after surface and temperature corrections.

In the present study, we make direct calculations of these thermal effects, and the ensemble averaged transport gap agrees excellently with experiment and previous estimations. The computational procedure can be summarized as follows: (1) Simulate the dynamic disorder induced by thermal fluctuation with *ab initio* Molecular Dynamics (AIMD). (2) Calculate the eigenvalues from density functional theory (DFT) on hundreds of disordered structures sampled from the MD trajectory. (3) Correct thousands of Kohn-Sham eigenvalues to GW-quasiparticle energies following the linear relations between the two quantities that were determined from rigorous GW calculations. The ensemble-averaged density of states provides us with a straightforward representation of the thermal effect on the near gap band structure.

## 2. Computational details

AIMD simulations within the Born-Oppenheimer ground state potential surface were performed with the CPMD package[22] on a primitive cell consisting two pentacene molecules for ~15ps, and on a 2×2×1 super-cell consisting 8 pentacene molecules for ~7ps. The lattice parameters[23] and initial molecular geometry being kept the same as in Ref. 14. Periodic boundary conditions were imposed to mimic an infinite bulk system. The system temperature was kept around 300K using Nosé-Hoover chain thermostat[24,25] within the canonical NVT ensemble. The Γ point was used in the Brillouin zone sampling. A time step of 0.5fs was used to ensure good control of the conserved quantities. The Perdew-Burke-Ernzerhof (PBE) functional was used to calculate the total electronic energy and forces with kinetic energy cutoff of 70 Ry. Troullier-Martins norm-conserving pseudopotentials[26] were employed to represent the core electrons and nuclei.

To find the linear relations between the GW quasi-particle energies and DFT eigenvalues, GW calculations were performed starting from the DFT-PBE calculations using the BerkeleyGW package[27] for the static structure and 6 disordered structures sampled from AIMD simulation of the primitive cell. Quasiparticle energies ($\varepsilon^{QP}$) of the near gap states were evaluated using a Γ-centered 4×4×2 k-mesh.

Then the linear relationships between $\varepsilon^{QP}$ and the Kohn-Sham eigenvalues from DFT-PBE ($\varepsilon^{DFT}$) for the highest occupied and lowest unoccupied states were determined and applied to correct hundreds of thousands of DFT-PBE eigenvalues extracted from the electronic structure calculations on hundreds of disordered structures sampled from AIMD on 2×2×1 super-cell. These electronic structure calculations as well as the DFT starting point for the GW calculations were all carried out using the Quantum Expresso package[28] with Troullier-Martins norm-conserving pseudopotentials and plane-wave basis functions with a 400eV cutoff.

## 3. Results and discussion

**3.1 MD simulation results analysis.** AIMD finds a thermal equilibrium structure essentially identical to the static structure for the pentacene crystal. The averages of the C-H and C-C bond lengths from AIMD are basically equal to the averages in the static structure respectively, as listed in Table S1. The former is obtained by averaging over hundreds of thousands of structures sampled by AIMD. The latter is just within one static geometry where the C-C bonds are slightly different from one to another, so are the C-H bonds. The thermal effects on the C-C and C-H bond length are manifested by the standard deviations of the bond lengths from AIMD simulations, i.e. 0.038 Å for C-C bond and 0.029 for the C-H bond, which also indicate mild thermal fluctuations in the local geometry of solid pentacene at room temperature.

To characterize its nonlocal geometry, the center of mass (COM) distance within four different pentacene pairs (shown in Figure 1) are specifically considered in the 2×2×1 super-cell. The COM distance within four molecular pairs at the initial static structure and their corresponding averages during the simulation are listed in Table S1. The average COM distances between molecules during the AIMD simulation are within 0.2% of the 0 K structure with deviations of 0.1-0.2 Å. Since the thermal equilibrium structure reproduces the 0K structure, AIMD is a valid way to predict the fluctuation of the ionic and electronic structure at finite temperature. This justifies the prediction of intermolecular distances from the AIMD simulation using 2×2×1 super-cell and indicates moderate thermal fluctuation in the intermolecular geometries during the AIMD simulations.

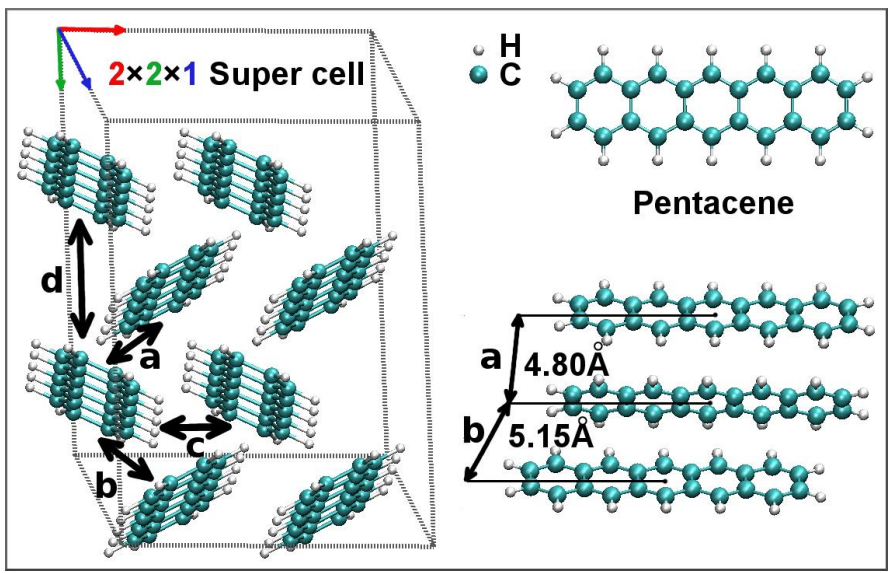

**Figure 1.** Pentacene molecule (top right) and crystal structure with a 2×2×1 super-cell (left) where the pentacene pairs considered in the geometry analysis in the contents are indicated with black thick arrows; the bottom right is an illustration of the structure of pair a and b with the respective center of mass distances.

To further understand the intermolecular interactions within the four types of pentacene pairs, we calculate the binding energies at B3LYP/6-311G** level with the Gaussian 09 package, and find that the a≥b>c≈d, agree with the previous reports in the literature[9,29]. Note that pair a and b in Figure 1 have similar structure but the COM distance of the former is shorter than that of the latter by 0.35 Å, which implies that the intermolecular interaction between the two molecules in pair a might be stronger than that in pair b. However, the binding energy of the isolated pentacene dimer in pair a is only 2 meV greater than that in pair b.

Above all, the AIMD will meaningfully express the thermal fluctuation about the 0K results previous published with an *ad hoc* broadening of the PES. The AIMD results do not support temperature dependent change in the equilibrium structure.

Further validation of the fluctuations produced by the AIMD simulation is provided via simulated vibrational spectroscopy of the pentacene crystal. The dipole moment of the entire box was computed within the maximally localized Wannier function scheme during the simulation[30,31]. Then, linear response theory was used to derive the infrared spectrum by doing a Fourier transform of the dipole-dipole auto-correlation function[32,33]. The good agreement between the calculated and experimental IR spectra (Figure

2) suggests that the molecular vibrations are properly characterized in the simulations and the resulting disordered systems are good sampling of what would be expected in the experiment. It is worth mentioning that the amplitudes of the molecular vibrations during the simulation are smaller than the amplitudes of the normal modes of the isolated pentacene molecule. We quantify this by calculating the root mean square deviation (RMSD) of each pentacene molecule in the super-cell from its equilibrium structure along the AIMD trajectory with respect to its initial static structure. The maximum, minimum and the average of the RMSD over all eight molecules, are 0.29 Å, 0.08 Å and 0.16±0.03 Å. In contrast, for an isolated 0K pentacene, using the normal mode coordinates of all the 102 vibrational modes of an optimized isolated pentacene and bose-einstein statistics, we find 0.61 Å, 0 Å and 0.12±0.11 Å respectively.

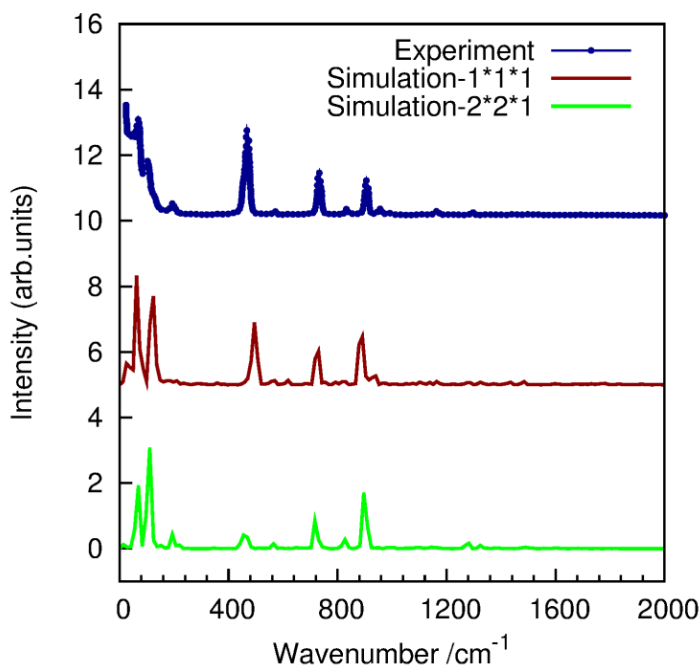

**Figure 2.** Comparison of the IR spectra from AIMD simulations of solid pentacene with a 1×1×1 primitive cell (dark-red) and a 2×2×1 super cell (green) to the high-resolution electron energy-loss spectrum of a multi-layer pencacene on Ag (111) (dark-blue) [Hendrik Adriaan van Laarhoven, PhD Thesis, *The Influence of the Vibrational Properties on Charge Transport in Oligoacenes,* Section 4.4.1, p43, Fig 4.9]. The quantitative agreement between the calculated and experimental IR spectra suggests that the molecular vibrations are properly characterized in the simulations and the resulting disordered systems are a good sampling of what would be expected in the experiment.

To investigate the effect of the intermolecular vibrations on the electronic structure, we calculated the Wannier orbitals and the Wannier centers[30] of the localized HOMO and HOMO-7 within the 2×2×1 super-cell along a section of the MD trajectory from 6.2 ps to 7.2 ps, as these two localized orbitals are related to the band extrema of the HOMO band. The two Wannier centers are located on two cofacial pentacene molecules, with the HOMO orbital centered on a central ring C-C π bond and the HOMO-7 orbital centered on a terminal ring C-C π bond, as shown in Figure 4. As the nuclear geometry changes, the Wannier orbitals of HOMO and HOMO-7 remain localized on qualitatively the same sites, but as the two molecules move during the simulation, these orbital centers move slightly with respect to their molecules by 0.25±0.12 Å and 0.35±0.12 Å respectively. The distance between the two Wannier centers fluctuate around 9.07 Å, while the center of mass distance between the two molecules where these Wannier centers are located fluctuates around 7.90 Å. Figure 3 also shows the comparison of the evolution of these two distances along the simulation time, where the former fluctuates in similar pattern as the latter, providing evidence for direct modulation of electronic structure by the change of intermolecular distance. However, since electronic motion in this simulation is essentially instantaneous (due to the Born Oppenheimer approximation) and coupled to higher frequency intramolecular vibrations as well as the intermolecular vibrations captured by the COM distances, the fluctuations of the Wannier centers are more complicated than that of the center of mass distances.

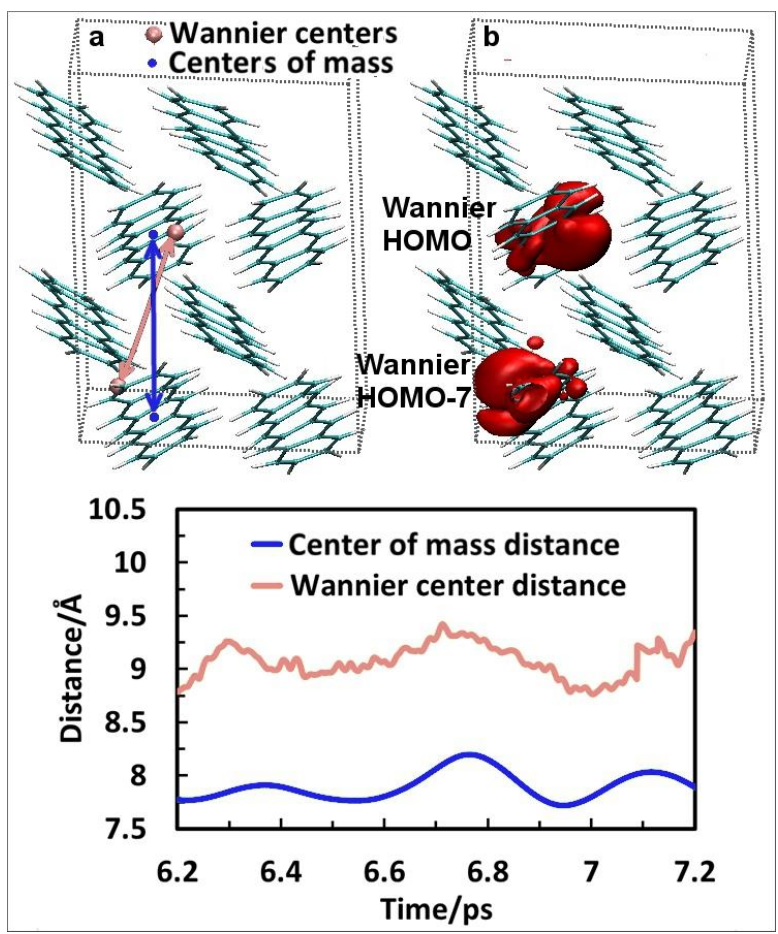

**Figure 3.** The Wannier centers (a/pink spheres) and the corresponding Wannier orbitals (b) of HOMO and HOMO-

7 within the 2×2×1 super-cell. The center of mass distance between the two molecules where the two Wannier center located are indicated by the blue dots and arrow in (a). The nuclear geometry is taken from the snapshot of the AIMD trajectory at 6.201 ps. The isosurface value of the orbitals is 0.02. The bottom plot is the evolution of the Wannier center distance between the localized HOMO and HOMO-7 (orange) and the center of mass distance between the related two pentacene molecules along the simulation time.

**3.2 Linear relationship between GW quasi-particle energies and DFT eigenvalues.**
The *ab initio* GW approach, as a method of first-principles many-body perturbation theory based on the Green's function (G), approximating the electron self-energy (Σ) as the first term of the expansion in the screened Coulomb interaction (W), has been successfully applied to the study of the quasiparticle (QP) property of organic solids, and has proven to yield quantitatively accurate quasiparticle band gaps and dispersion relations from first-principles.[14,27,34,35] In practice, the quasi-particle energy ($\varepsilon^{QP}$) within the GW approximation is evaluated in the form of Dyson's equation, in the basis of the Kohn-Sham orbitals through the computation of the diagonal and off-diagonal elements of $\Sigma$[27]. It is often the case that DFT and QP wavefunctions are nearly the same, so the QP correction can be evaluated by Eq. (1) where $V_{xc}$ is the exchange-correlation potential of the stating DFT, $\psi_i$ is the Kohn-Sham orbital.

$$\varepsilon_i^{QP} - \varepsilon_i^{DFT} = \langle \psi_i | \Sigma - V_{xc} | \psi_i \rangle \quad (1)$$

It has been observed that there is a linear relation between the GW quasi-particle energies and DFT eigenvalues for states near the band gap $\varepsilon^{QP} - \varepsilon^{DFT} = s \cdot \varepsilon^{DFT} + \delta$[14,35,36], which implies that GW-corrected band structure and DOS can be obtained easily from the above linear relation without the need for explicit evaluation of GW corrections for large-scale calculations[14]. In the present work, we are interested in applying this linear relation broadly on hundreds of disordered structures to eventually obtain an ensemble average of GW-corrected density of states. So, we have revisited the GW calculations on the static structure, and found that for the highest occupied states: $\varepsilon^{QP} - \varepsilon^{DFT} = 0.3 \cdot \varepsilon^{DFT} - 0.6$ eV, while for the lowest unoccupied state: $\varepsilon^{QP} - \varepsilon^{DFT} = 0.3 \cdot \varepsilon^{DFT} + 0.6$ eV.

Since the plane wave basis functions used here are independent of nuclei position R and the QP correction in Eq.(1) consists largely of a constant shift with a weak dependence on R [35], By taking derivatives from both side of Eq.(1) with respect to the nuclei positions R, it becomes:

$$\frac{\partial(\varepsilon_i^{QP} - \varepsilon_i^{DFT})}{\partial R} = \left\langle \Psi_i \left| \frac{\partial(\Sigma - V_{xc})}{\partial R} \right| \Psi_i \right\rangle + 2 \left\langle \frac{\Psi_i}{\partial R} \left| \Sigma - V_{xc} \right| \Psi_i \right\rangle$$

$$= \left\langle \Psi_i \left| \frac{\partial(\Sigma - V_{xc})}{\partial R} \right| \Psi_i \right\rangle \cong 0 \quad (2)$$

This suggests that the QP correction to the DFT eigenvalue is approximately independent of moderate geometric perturbations. To numerically verify this approximation, we carry out rigorous GW calculations on 6 disordered structures taken from the AIMD simulation performed on the primitive cell. The instantaneous structures were chosen carefully such that their potential energies spread through the range of thermal fluctuation, and time intervals of ~1ps were taken to de-correlate the successive structures. By plotting the $\varepsilon^{QP}$ against the corresponding $\varepsilon^{DFT}$ for a few eV of near gap states at each k-point from all 6 disordered structures, in Figure 4 we show that, the linear relationship between $\varepsilon^{QP}$ and $\varepsilon^{DFT}$ determined with the static structure holds in the disordered structures. However, the geometry fluctuations change each energy level with respect to the static structure, contributing to the broadening in the spectrum, as shown by the differences in the linear distributions of the red circles and the black triangles in Figure 4.

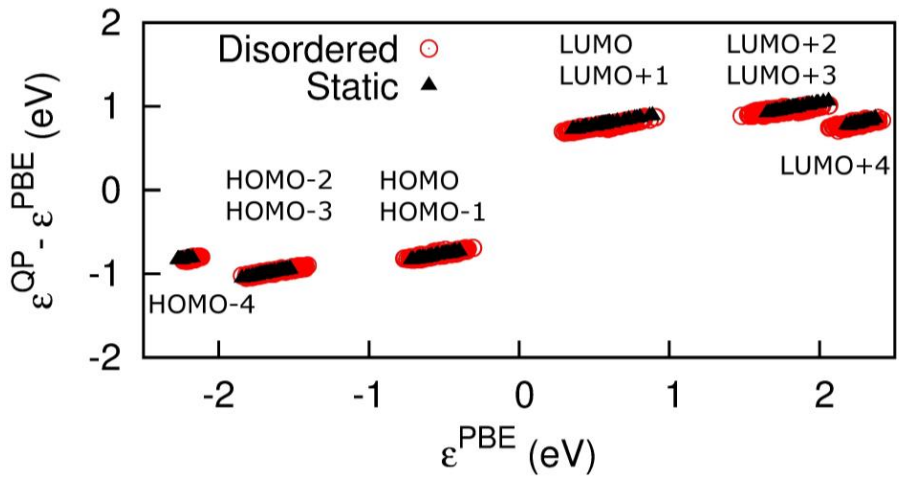

**Figure 4.** Linear relations between GW correction and the corresponding DFT-PBE eigenvalues for the five highest occupied and five lowest unoccupied states near the band gap. Zero is taken to be the middle of the gap.

Although for illustrative purpose, it's feasible to use a primitive cell with appropriate k-mesh applied to show that the linear relationship between the QP corrections and DFT eigenvalues is conserved at the degree of perturbation caused by heating to room temperature in AIMD, in practice, to calculate the band structure of solid pentacene with dynamic disorders, a super-cell should be used to capture the nature of the intermolecular interactions, which will characterize the disorders more completely than using a primitive cell with a k-mesh. This is due to the addition of artificial periodicity from the nuclei to the dynamic disorder. For pentacene in the pristine crystalline phase, it is shown by S. Sharifzadeh *et al*.[14] that using a Γ-centered k-point mesh of 2×2×1 and wave function basis cutoff of 400 eV is sufficient to converge the total DFT-PBE energy to 1 meV/atom. It is expected that by using 2×2×1 super-cell, the band extrema differences at Γ point will be good approximations for the band gap and band width.

**3.3 Approximate the band gap and band width from super-cell calculation.** Before applying the GW linear corrections to the large numbers of disordered structures sampled from AIMD, we checked the accuracy of the band structure calculation using a 2×2×1 super-cell for the disordered system. As illustrated in great detail elsewhere, effective band structure can be obtained by super-cell calculations in perturbed organic and inorganic solid state materials[38,39]. To justify the size of the 2×2×1 model system for solid pentacene in the presence of dynamic disorder concerned here, we took 4 disordered structures from the MD trajectory, carried out electronic structure calculations with a 2×2×2 k-mesh as well as at just the Γ point, compared the band gap, HOMO and LUMO band width from the former with the corresponding band extrema differences from the latter. We found that the errors in approximating the band gap and band width with the band extrema differences calculated using 2×2×1 super-cell at the Γ point are ~0.01eV (Table S2), which is much smaller than the standard deviation (0.026–0.031 eV) of the ensemble average of each energy level, thus longer range finite disorder effects would be washed out by the thermal fluctuation.

**3.4 Ensemble-averaged density of states.** 640 instantaneous structures with dynamic disorders were sampled from the last 2 ps of the AIMD simulation on the 2×2×1 super-cell. Then DFT-PBE electronic structure calculations were carried out on each of the disordered structures at the Γ point, DFT eigenvalues were extracted and corrected to quasiparticle energies following the linear equations determined above. The energy distribution of the GW-corrected eigenvalues of each near gap state from all of sampled disordered structures at the Γ point is shown in Figure 5.

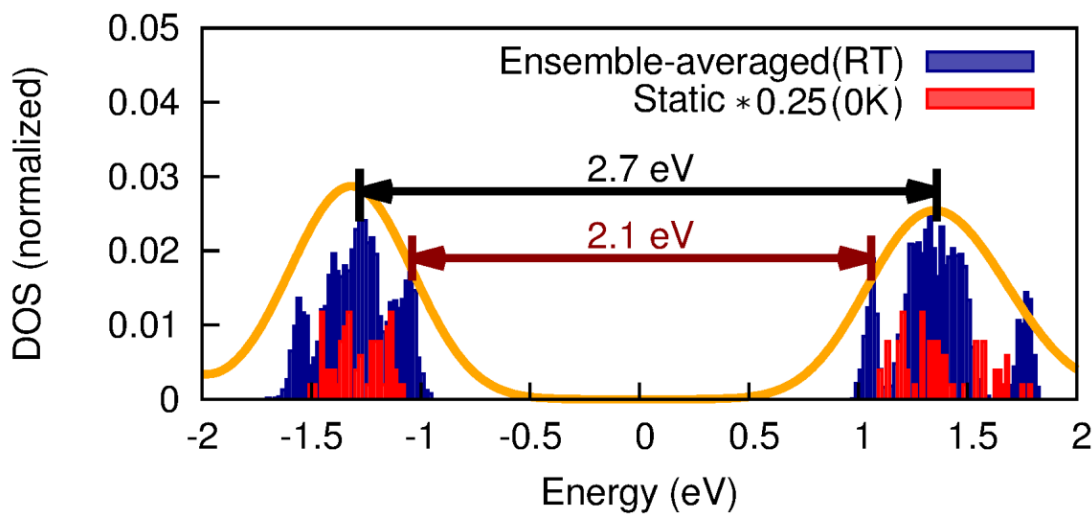

**Figure 5.** Normalized density of states (blue) from 640 disordered structures sampled from the AIMD simulation with 2×2×1 super-cell. The energy distribution is normalized within the sampling space. The normalized 0 K GW calculation (red) is a quarter of its original value for ease of comparison. The orange line is the 0 K GW spectrum convoluted with Gaussian functions with a width of 0.25 eV.

By taking the difference between the ensemble average of the highest occupied and lowest unoccupied levels, the edge-to-edge transport gap is found to be 2.1±0.04 eV with standard deviations of 0.026 eV from both sides. Comparing it with the GW transport gap from static structure calculation (2.2 eV in Table. 1) using PBE as DFT starting point as well, we attribute this 0.1 eV deviation to the intramolecular geometry change due to the extra potential energy gained when heating up the system to room temperature during the MD simulation, as the lattice vectors were fixed to experiment to circumvent the poor performance of PBE in describing the van der Waals interactions that is necessary for characterization of intermolecular spacing in weakly bound crystals[38,40,41]. The previous GW stactic calculation[14] using PBE as starting point agreed with the experimental edge-to-edge gap of 2.2 eV. This experimental edge-to-edge gap reported in Ref. 14 was estimated from the photoemission spectrum of Ref. 19. The difference between the current and previous results for the edge-to-edge gap also verify the estimation of ~0.1–0.2eV in shifting the photoemission spectrum, which was made to account for the fact that a static 0 K calculation was compared to finite-temperature measurements[14,18].

Although Γ point calculation leaves large spacing in the spectrum, with contributions from hundreds of disordered structures added up together, it still shows roughly a Gaussian profile in the density of the states on both sides of the band gap (Figure 5). This can be taken as a reoccurrence of the Boltzmann distribution of the states of a system governed by the thermal fluctuation, as the disordered structures were sampled with an equal time interval of 1.5fs along the AIMD trajectory. We take the difference between the peaks of the HOMO band and LUMO band as the peak-to-peak band gap, which is 2.7 eV (Figure 5). When Gaussian broadening with width of 0.25eV was applied to the GW spectrum using PBE starting point at 0 K, to account for the finite temperature effect, the peak-to-peak gap is also found to be ~2.7 eV, as shown by the orange curve in Figure 5. The spectrum convoluted with uniform Gaussians does approximate the ensemble-averaged spectrum to some extent, but doesn't necessarily capture the details of the spectrum, especially for the HOMO band.

The comparison of the peak-to-peak gaps from the AIMD simulations at room temperature, GW static calculations at 0 K and the experimental Photoemission spectra is given in Table 1. The GW-corrected ensemble average at room temperature of the peak-to-peak gap is in very good agreement with the experimental value 2.7 eV after corrections of 0.5 eV for the surface and 0.2 eV for the Frank-Condon effects[18].

**Table 1.** Comparison of the edge-to-edge and peak-to-peak gaps, in eV from AIMD simulations at room temperature (RT), GW static calculations at 0 K and Photoemission spectra

|  | GW-corrected Ensemble average (RT) | GW Static (0 K) | Experiment Photoemission |
| --- | --- | --- | --- |
| Edge-to-edge gap | 2.1±0.04 | 2.2 | 2.2* |
| Peak-to-peak gap | 2.7 | 2.7 | 2.7[18] |

*Taken from Ref. 14, which was estimated from photoemission spectrum of Ref.18

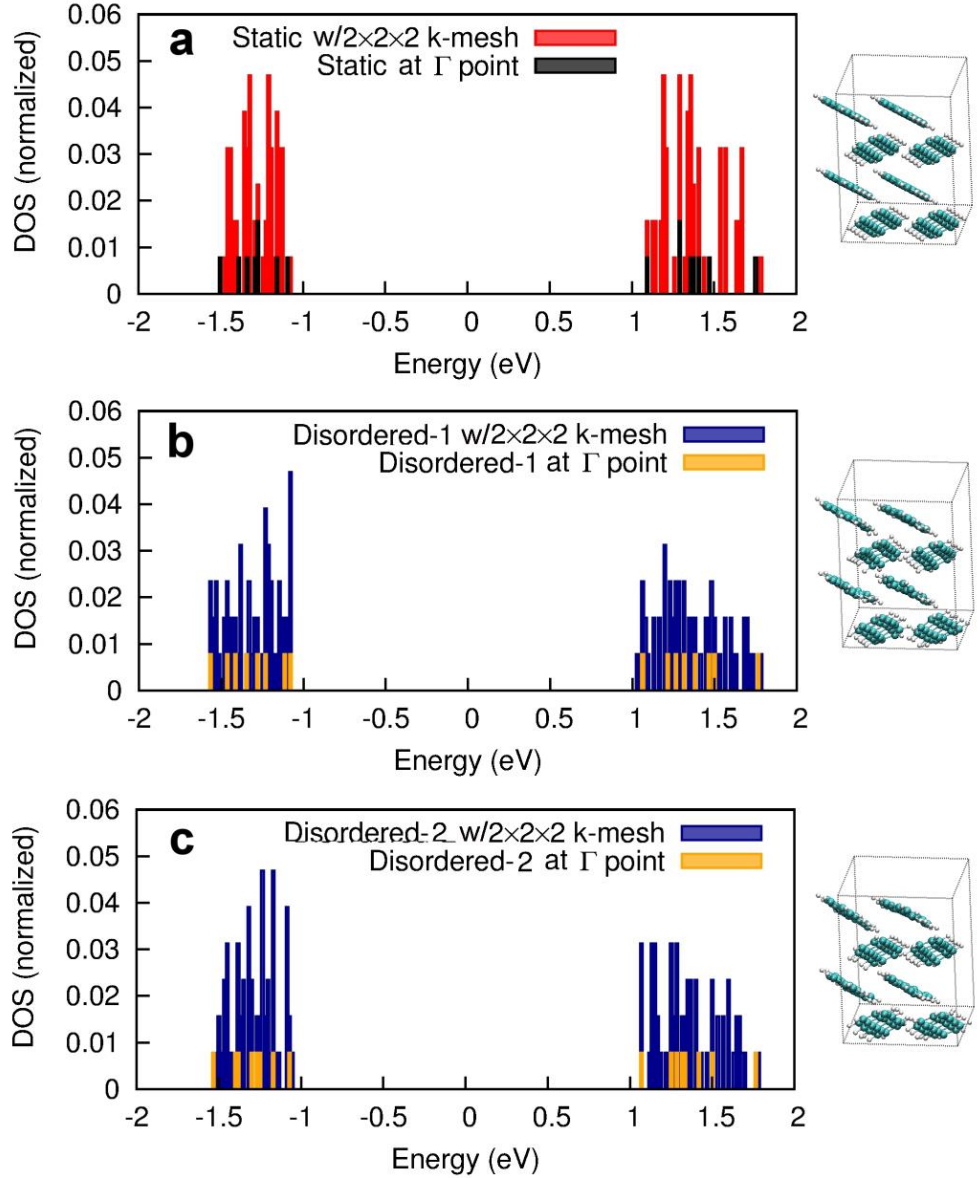

**Figure 6.** Normalized Density of states (DOS) for the static (a) and two disordered structures (b and c) at Γ point (black for the static structure, orange for the disordered structures) * 0.25 and with 2×2×2 k-mesh (red for the static structure, blue for the disordered structures).

It is also noticed that the spectrum has a three-peak structure in both side of the band gap. We compared the DOS using 2×2×2 k-mesh and that at Γ point for the static structure and two of the disordered structures mentioned in section 3.3, in Figure 6 it shows that, for both the static and disordered structures, the band extrema at Γ point calculation approximate the band edges very well. Although Γ point calculation leave large space in the spectra, it was able to capture the states where the peaks are in the spectra using 2×2×2 k-mesh for most of the cases. However, this is likely to lead to the obvious three peak structure in the ensemble averaged spectrum shown in Figure 5. We expect that it would have less of the three-peak structure with more k-points used in the calculations, notwithstanding the possibility that the three peak structure came from structure like in Figure 6b and was an intrinsic detail lost in experiment, especially for the HOMO band. In the HOMO band of the spectrum at Γ point in Figure 6b, the near equally distributed states in the middle do not correspond to the strongest peak of the spectrum in blue. This suggests that to obtain a more accurate profile of the spectrum requires large scale calculations with sufficient k-point sampling or improvement of the resolution in experimental measurements.

## 4. Conclusions

In conclusion, the ensemble-averaged density of states was obtained by combining GW approach and AIMD simulations for solid pentacene. The linear relations between GW quasi-particle correction and the DFT-PBE eigenvalues were shown to be conserved at the degree of perturbation caused by heating to room temperature in AIMD. The linear relations were then applied on hundreds of disordered structures sampled from AIMD. We find the edge-to-edge transport gap including thermal effect is 2.1±0.04 eV, reduced by ~ 0.1 eV in contrast to the static structure. The peak-to-peak gap is found to be 2.7eV, providing fully *ab initio* agreement with experiment where previous theory required *ad hoc* Gaussian broadening and temperature corrections.

**Acknowledgements**

Acknowledgements are made to Molecular Foundry at Lawrence Berkeley National Laboratory, supported by the Office of Science, Office of Basic Energy Sciences, of the U.S. Department of Energy. This research used computational resources of the National Energy Research Scientific Computing Center, which is supported by the Office of Science of the US Department of Energy.

**References**

(1) Rogers, J. A.; Someya, T.; Huang, Y. Materials and Mechanics for Stretchable Electronics. *Science* **2010**, *327* (5973), 1603–1607.

(2) Sanvito, S. Organic Electronics: Spintronics Goes Plastic. *Nat. Mater.* **2007**, *6* (11), 803–804.

(3)  Vardeny, Z. V.; Heeger, A. J.; Dodabalapur, A. Fundamental Research Needs in Organic Electronic Materials. *Synth. Met.* **2005**, *148* (1), 1–3.

(4)  Kippelen, B.; Brédas, J.-L. Organic Photovoltaics. *Energy Environ. Sci.* **2009**, *2* (3), 251.

(5)  Forrest, S. R. The Path to Ubiquitous and Low-Cost Organic Electronic Appliances on Plastic. *Nature* **2004**, *428* (6986), 911–918.

(6)  Lu, L.; Kelly, M. A.; You, W.; Yu, L. Status and Prospects for Ternary Organic Photovoltaics. *Nat. Photonics* **2015**, *9* (8), 491–500.

(7)  Difley, S.; Wang, L.-P.; Yeganeh, S.; Yost, S. R.; Van Voorhis, T. Electronic Properties of Disordered Organic Semiconductors via QM/MM Simulations. *Acc. Chem. Res.* **2010**, *43* (7), 995–1004.

(8)  Coropceanu, V.; Cornil, J.; da Silva Filho, D. A.; Olivier, Y.; Silbey, R.; Brédas, J.-L. Charge Transport in Organic Semiconductors. *Chem. Rev.* **2007**, *107* (4), 926–952.

(9)  Troisi, A.; Orlandi, G. Charge-Transport Regime of Crystalline Organic Semiconductors: Diffusion Limited by Thermal Off-Diagonal Electronic Disorder. *Phys. Rev. Lett.* **2006**, *96* (8), 86601.

(10) Refaely-Abramson, S.; Jain, M.; Sharifzadeh, S.; Neaton, J. B.; Kronik, L. Solid-State Optical Absorption from Optimally Tuned Time-Dependent Range-Separated Hybrid Density Functional Theory. *Phys. Rev. B* **2015**, *92* (8), 81204.

(11) Coto, P. B.; Sharifzadeh, S.; Neaton, J. B.; Thoss, M. Low-Lying Electronic Excited States of Pentacene Oligomers: A Comparative Electronic Structure Study in the Context of Singlet Fission. *J. Chem. Theory Comput.* **2015**, *11* (1), 147–156.

(12) Sharifzadeh, S.; Wong, C. Y.; Wu, H.; Cotts, B. L.; Kronik, L.; Ginsberg, N. S.; Neaton, J. B. Relating the Physical Structure and Optoelectronic Function of Crystalline TIPS-Pentacene. *Adv. Funct. Mater.* **2015**, *25* (13), 2038–2046.

(13) Sharifzadeh, S.; Darancet, P.; Kronik, L.; Neaton, J. B. Low-Energy Charge-Transfer Excitons in Organic Solids from First-Principles: The Case of Pentacene. *J. Phys. Chem. Lett.* **2013**, *4* (13), 2197–2201.

(14) Sharifzadeh, S.; Biller, A.; Kronik, L.; Neaton, J. B. Quasiparticle and Optical Spectroscopy of the Organic Semiconductors Pentacene and PTCDA from First Principles. *Phys. Rev. B* **2012**, *85* (12), 125307.

(15) Ueno, N.; Kera, S. Electron Spectroscopy of Functional Organic Thin Films: Deep Insights into Valence Electronic Structure in Relation to Charge Transport Property. *Prog. Surf. Sci.* **2008**, *83* (10–12), 490–557.

(16) Zahn, D. R. T.; Gavrila, G. N.; Gorgoi, M. The Transport Gap of Organic Semiconductors Studied Using the Combination of Direct and Inverse Photoemission. *Chem. Phys.* **2006**, *325* (1), 99–112.

(17) Watkins, N. J.; Zorba, S.; Gao, Y. Interface Formation of Pentacene on Al[sub 2]O[sub 3]. *J. Appl. Phys.* **2004**, *96* (1), 425.

(18) Amy, F.; Chan, C.; Kahn, A. Polarization at the Gold/pentacene Interface. *Org. Electron.* **2005**, *6* (2), 85–91.

(19) Fukagawa, H.; Yamane, H.; Kataoka, T.; Kera, S.; Nakamura, M.; Kudo, K.; Ueno, N. Origin of the Highest Occupied Band Position in Pentacene Films from Ultraviolet Photoelectron Spectroscopy: Hole Stabilization versus Band Dispersion. *Phys. Rev. B* **2006**, *73* (24), 245310.

(20) Hill, I. G.; Kahn, A.; Soos, Z. G.; Pascal, Jr, R. A. Charge-Separation Energy in Films of π-Conjugated Organic Molecules. *Chem. Phys. Lett.* **2000**, *327* (3–4), 181–188.

(21) Kakuta, H.; Hirahara, T.; Matsuda, I.; Nagao, T.; Hasegawa, S.; Ueno, N.; Sakamoto, K. Electronic Structures of the Highest Occupied Molecular Orbital Bands of a Pentacene Ultrathin Film. *Phys. Rev. Lett.* **2007**, *98* (24), 247601.

(22) Dominik Marx and Jürg Hutter. Ab Initio Molecular Dynamics: Theory and Implementation. In *Modern Methods and Algorithms of Quantum Chemistry, Proceedings*; Grotendorst, J., Ed.; John von Neumann Institute for Computing, 2000; pp 329–477.

(23) Lattice Parameters.

(24) Hoover, W. Canonical Dynamics: Equilibrium Phase-Space Distributions. *Phys. Rev. A* **1985**, *31* (3), 1695–1697.

(25) Nosé, S. A Unified Formulation of the Constant Temperature Molecular Dynamics Methods. *J. Chem. Phys.* **1984**, *81* (1), 511.

(26) Troullier, N.; Martins, J. L. Efficient Pseudopotentials for Plane-Wave Calculations. *Phys. Rev. B* **1991**, *43* (3), 1993–2006.

(27) Deslippe, J.; Samsonidze, G.; Strubbe, D. A.; Jain, M.; Cohen, M. L.; Louie, S. G. BerkeleyGW:

A Massively Parallel Computer Package for the Calculation of the Quasiparticle and Optical Properties of Materials and Nanostructures. *Comput. Phys. Commun.* **2012**, *183* (6), 1269–1289.

(28) Giannozzi, P.; Baroni, S.; Bonini, N.; Calandra, M.; Car, R.; Cavazzoni, C.; Ceresoli, D.; Chiarotti, G. L.; Cococcioni, M.; Dabo, I.; Dal Corso, A.; de Gironcoli, S.; Fabris, S.; Fratesi, G.; Gebauer, R.; Gerstmann, U.; Gougoussis, C.; Kokalj, A.; Lazzeri, M.; Martin-Samos, L.; Marzari, N.; Mauri, F.; Mazzarello, R.; Paolini, S.; Pasquarello, A.; Paulatto, L.; Sbraccia, C.; Scandolo, S.; Sclauzero, G.; Seitsonen, A. P.; Smogunov, A.; Umari, P.; Wentzcovitch, R. M. QUANTUM ESPRESSO: A Modular and Open-Source Software Project for Quantum Simulations of Materials. *J. Phys. Condens. Matter* **2009**, *21* (39), 395502.

(29) Sánchez-Carrera, R. S.; Paramonov, P.; Day, G. M.; Coropceanu, V.; Brédas, J.-L. Interaction of Charge Carriers with Lattice Vibrations in Oligoacene Crystals from Naphthalene to Pentacene. *J. Am. Chem. Soc.* **2010**, *132* (41), 14437–14446.

(30) Marzari, N.; Mostofi, A. A.; Yates, J. R.; Souza, I.; Vanderbilt, D. Maximally Localized Wannier Functions: Theory and Applications. *Rev. Mod. Phys.* **2012**, *84* (4), 1419–1475.

(31) Marzari, N.; Vanderbilt, D. Maximally Localized Generalized Wannier Functions for Composite Energy Bands. *Phys. Rev. B* **1997**, *56* (20), 12847–12865.

(32) Berens, P. H. Molecular Dynamics and Spectra. I. Diatomic Rotation and Vibration. *J. Chem. Phys.* **1981**, *74* (9), 4872.

(33) Mathias, G.; Ivanov, S. D.; Witt, A.; Baer, M. D.; Marx, D. Infrared Spectroscopy of Fluxional Molecules from (Ab Initio) Molecular Dynamics: Resolving Large-Amplitude Motion, Multiple Conformations, and Permutational Symmetries. *J. Chem. Theory Comput.* **2012**, *8* (1), 224–234.

(34) Rohlfing, M.; Louie, S. G. Electron-Hole Excitations and Optical Spectra from First Principles. *Phys. Rev. B* **2000**, *62* (8), 4927–4944.

(35) Hybertsen, M. S.; Louie, S. G. Electron Correlation in Semiconductors and Insulators: Band Gaps and Quasiparticle Energies. *Phys. Rev. B* **1986**, *34* (8), 5390–5413.

(36) Kümmel, S.; Kronik, L. Orbital-Dependent Density Functionals: Theory and Applications. *Rev. Mod. Phys.* **2008**, *80* (1), 3–60.

(37) Ismail-Beigi, S.; Louie, S. G. Excited-State Forces within a First-Principles Green's Function Formalism. *Phys. Rev. Lett.* **2003**, *90* (7), 76401.

(38) Medeiros, P. V. C.; Stafström, S.; Björk, J. Effects of Extrinsic and Intrinsic Perturbations on the

Electronic Structure of Graphene: Retaining an Effective Primitive Cell Band Structure by Band Unfolding. *Phys. Rev. B* **2014**, *89* (4), 41407.

(39) Popescu, V.; Zunger, A. Extracting E versus K⃗ Effective Band Structure from Supercell Calculations on Alloys and Impurities. *Phys. Rev. B* **2012**, *85* (8), 85201.

(40) Tkatchenko, A.; Romaner, L.; Hofmann, O. T.; Zojer, E.; Ambrosch-Draxl, C.; Scheffler, M. Van Der Waals Interactions Between Organic Adsorbates and at Organic/Inorganic Interfaces. *MRS Bull.* **2011**, *35* (6), 435–442.

(41) Nabok, D.; Puschnig, P.; Ambrosch-Draxl, C. Cohesive and Surface Energies of π-Conjugated Organic Molecular Crystals: A First-Principles Study. *Phys. Rev. B* **2008**, *77* (24), 245316.
16

**Supporting Information**

**Exploring the Influence of Dynamic Disorder on Transport Gap in Solid Pentacene**

Zhiping Wang,[1,†] Sahar Sharifzadeh,[1,‡] Zhenfei Liu[1,¶], Peter Doak[1,§] and Jeffrey B. Neaton[1,2,3]

[1]*Molecular Foundry, Lawrence Berkeley National Laboratory, Berkeley, California 94720, United States*

[2]*Department of Physics, University of California Berkeley, Berkeley, California 94720, United States*

[3]*Kavli Energy NanoScience Institute at Berkeley, Berkeley, California 94720, United States*

[†] Present address: School of Physics, Shandong University, Jinan 250100, P. R. China

[‡] Present address: Department of Electrical and Computer Engineering, Boston University, Boston, MA 02215, United States

[¶] Present address: Department of Chemistry, Wayne State University, Detroit, MI 48202, United States

[§] Present address: Center for Nanophase Material Science, Oakridge National Laboratory, Oakridge Tennessee 37831, United States

**Table S1.** Comparison of the ensemble averages of the C-H, C-C bond length and the center of mass distance between pentacene molecules in pair a, b, c and d (defined in Figure 1.) from the AIMD simulation using 2×2×1 super-cell with those from the static structure (all in Angstrom).

|  | Ensemble average (RT) | Static (0 K) |
|---|---|---|
| C-H bond length | 1.10±0.03 | 1.09±0.01 |
| C-C bond length | 1.41±0.04 | 1.41±0.03 |
| COM/pair a | 4.80±0.12 | 4.80 |
| COM/pair b | 5.16±0.10 | 5.15 |
| COM/pair c | 6.06±0.11 | 6.06 |
| COM/pair d | 7.90±0.10 | 7.90 |

**Table S2.** Comparison of band extrema differences at $\Gamma$ point using 2×2×1 super-cell with the band width and band gap calculated by using 2×2×1 super-cell with a 2×2×2 k-mesh for both of the static and disordered structures. Four disordered structures are sampled from the AIMD trajectory with intervals of ~500fs, which are numbered as Disordered 1, 2, 3 and 4.

|  | HOMO band | | LUMO band | | Transport gap | |
|---|---|---|---|---|---|---|
|  | k=$\Gamma$ | k=2×2×2 | k=$\Gamma$ | k=2×2×2 | k=$\Gamma$ | k=2×2×2 |
| Static | 0.32 | 0.32 | 0.51 | 0.53 | 0.75 | 0.75 |
| Disordered 1 | 0.37 | 0.36 | 0.55 | 0.57 | 0.72 | 0.70 |
| Disordered 2 | 0.36 | 0.36 | 0.53 | 0.55 | 0.72 | 0.71 |
| Disordered 3 | 0.40 | 0.40 | 0.54 | 0.55 | 0.71 | 0.71 |
| Disordered 4 | 0.38 | 0.38 | 0.52 | 0.54 | 0.70 | 0.69 |